\documentclass[aps,prd,preprint,superscriptaddress,showpacs,nofootinbib,epsf]{revtex4}

\usepackage{graphicx}

\begin{document}


\title{Gravitational Radiation \\ from Axisymmetric Rotational Core Collapse}


\author{Kei Kotake}
\email[E-mail: ]{kkotake@utap.phys.s.u-tokyo.ac.jp}
\affiliation{Department of Physics, School of Science, the University of
Tokyo, 7-3-1 Hongo, Bunkyo-ku, Tokyo 113-0033, Japan}
\author{Shoichi Yamada}
\affiliation{Science \& Engineering, Waseda University, 3-4-1 Okubo, Shinjuku,
Tokyo, 169-8555, Japan}
\author{Katsuhiko Sato}
\affiliation{Department of Physics, School of Science, the University of
Tokyo, 7-3-1 Hongo, Bunkyo-ku, Tokyo 113-0033, Japan}
\affiliation{Research Center for the Early Universe, School of Science,
the University of Tokyo, 7-3-1 Hongo, Bunkyo-ku, Tokyo 113-0033, Japan}

\date{\today}

\begin{abstract}
We have done a series of two-dimensional hydrodynamic simulations of the
rotational collapse of a supernova core in axisymmetry. We have employed
 a realistic equation of state (EOS) and taken into account electron
 captures and neutrino transport by the so-called leakage scheme. It is
 an important progress to apply the realistic EOS coupled with the
 microphysics to 2-D simulations for computing gravitational radiation in
 rotational core collapse.
 We have used the
quadrupole formula to calculate the amplitudes and the waveforms of
gravitational wave assuming the Newtonian gravity. 
From these
computations,
we have extended the conventional category of the gravitational waveforms.
Our results have shown that the peak amplitudes of
gravitational wave are mostly within the
 sensitivity range of the laser interferometers such as TAMA and first
 LIGO for a source at a distance of 10 kpc. 
Furthermore we have found that the amplitudes of the second peaks 
are within the detection limit of first LIGO for the source and 
first pointed out the
importance of the detections, since they will give us
the information as to the angular momentum distribution of evolved
massive stars. 
\end{abstract}


\maketitle

\section{Introduction}
Asymmetric core collapse and supernova have been supposed to be one of
the most plausible source of gravitational radiation for the 
long-baseline laser interferometers (GEO600, LIGO, TAMA,
VIRGO) ~\cite{thorne}. The detection of the gravitational signal is
important not only for the direct confirmation of general relativity but
also for the understanding of supernovae themselves because the gravitational
wave is only a window that enables us to see
directly the innermost part of an evolved star, where the angular
momentum distribution and the equation of state are unknown. 

Observationally, the asymmetric aspects of the dynamics of supernovae
are evident because they are confirmed by many observations of SN 1987A
\cite{cropper,papa,wang} and by the kick velocity of pulsars \cite{lyne}.
On the other hand, there is no consensus of the origin of asymmetry
theoretically. However, 
provided the facts that the progenitors of collapse-driven supernovae are a rapid rotator on the main sequence \cite{tass} and that the recent
theoretical studies suggest a fast rotating core prior to the collapse
\cite{heger00}, rotation should play an important role as the origin of
the asymmetric motions in the core collapse. It is noted that anisotropic neutrino radiations induced by 
rotation \cite{kotake} could induce a jet-like explosion
\cite{shimi01} as suggested by the observations of SN 1987A.  

So far there have been works devoted
to study the gravitational radiation in the rotational core
collapse \cite{mm,ys,zweg,rampp,dimmel,fry,shibata} (see \cite{new} for a 
review). To say rigorously, reliable core collapse simulations should
require the implementation of a realistic equation of state  (EOS), an adequate
treatment of microphysics (electron captures and other weak interactions) and neutrino transport, and a relativistic treatment of
gravitation. However it is difficult to incorporate all of them at the
same time. Therefore previous investigations have neglected or
approximated the above requirements partially. So far most of the 
computations have oversimplified the microphysics. In
addition, it has been reported that
general relativity does
not alter the significant features of gravitational radiation compared with those obtained in Newtonian
approximation such as the range of gravitational wave amplitudes and
frequencies \cite{dimmel}. This situation motivates us to employ a realistic EOS and treat
microphysics adequately in the Newtonian gravity. 
In this paper, we will study the wave forms of gravitational radiation
 elaborately by performing the
improved rotational core collapse simulations and will discuss what
information can be extracted from the analysis.

We describe the numerical models in the next section. In the third
section, we show the main numerical results. Conclusion is 
given in the last section.

\section{Models and Numerical Methods}
\subsection{Initial models}
We know little of angular momentum distributions in a core of evolved massive stars. Although it is supposed that some instabilities grow and transport angular momentum during the quasi-static evolutions, which mode prevails in what time scale is not understood very well at present. Therefore, we assume in this study 
the following two possible rotation laws. 

1. shellular rotation:
\begin{equation}
\Omega(r) = \Omega_0 \times \frac{{R_{0}}^2}{r^2 + R_{0}^2},
\end{equation}
where $\Omega(r)$ is an angular velocity, $r$ is a radius, and $\Omega_0, R_{0}$ are model constants. 

2. cylindrical rotation:
\begin{equation}
\Omega(X,Z) = \Omega_0 \times \frac{{X_{0}}^2}{X^2 + {X_{0}}^2} \cdot
\frac{Z_{0}^4}{Z^4 + Z_{0}^4},
\end{equation}
where $X$ and $Z$ denote distances from the rotational axis and the equatorial plane and $X_0, Z_0$ are model constants. The other parameters have the same meanings as above.

\begin{table}
\caption{The Model Parameters.}
\label{table:1}
\begin{center}
\begin{tabular}{ccccc} \hline \hline
Model & $T/|W|_{\rm init} (\%) $   & Rotation Law &
 $ R_{0}$, $X_{0}$, $Z_0$ $\times 10^8$ (cm) & $\Omega_0\,\,(\rm{s}^{-1})$   \\ \hline

 SSL & 0.25          & Shellular        &  $R_0 = 1 $  & 2.8   \\ 
 SSS & 0.25          & Shellular        &  $R_0 = 0.1 $  & 45.1 \\ 
 SCL  & 0.25          & Cylinder        &  $X_0 = 1 $, $Z_0 =1 $ & 2.7
	\\ 
 SCS  & 0.25          & Cylinder        &  $X_0 = 0.1$, $Z_0 =1 $ & 31.3\\    
 MSL & 0.50          & Shellular        &  $R_0 = 1 $  & 4.0   \\
 MSS & 0.50          & Shellular        &  $R_0 = 0.1$  & 63.4  \\
 MCL  & 0.50          & Cylinder        &  $X_0 = 1$, $Z_0 =1 $ & 3.8   \\  
 MCS  & 0.50          & Cylinder        &  $X_0 = 0.1$, $Z_0 =1 $ & 44.4  \\
 RSL & 1.50          & Shellular        &  $R_0 = 1$  & 6.8  \\
 RSS & 1.50          & Shellular        &  $R_0 = 0.1$  & 112  \\
 RCL  & 1.50          & Cylinder        &  $X_0 = 1$, $Z_0 =1 $ & 6.6    \\
 RCS  & 1.50          & Cylinder        &  $X_0 = 0.1$, $Z_0 =1$ & 76.8 \\\hline \hline
\end{tabular}
\end{center}
\end{table}

 Although recent theoretical studies \cite{heger00} give
 estimates for the angular velocity prior to the collapse, they are
 one-dimensional models with uncertainties and not the final
 answer. Hence we prefer a parametric approach in this paper. We have
 computed twelve models changing the combination of the total angular
 momentum, the rotation law, and the degree of differential
 rotation. The model parameters are presented in Table
 \ref{table:1}. The models are named after this combination, with the
 first letter ,``S (Slow)'', ``M (Moderate)'', ``R (Rapid)'',
 representing the initial $T/|W|_{\rm init}$, the second letter, ``S (Shellular),C
 (Cylindrical)'', denoting the rotation law, and the third letter ,``L
 (Long), S(Short)'', indicating the values of $R_{0}$ and $X_{0}$, which
 represent the degree of differential rotation. The initial ratio of
 rotational energy to gravitational energy is designated as $T/|W|_{\rm init}$. We
 have chosen $0.25, 0.5, 1.5 \%$ for $T/|W|_{\rm init}$. The rotational
 progenitor model \cite{rf:fry00} corresponds to Model MSL (Moderate,
 Shellular rotation, $R_0$ = 1000 km) in our simulations and we take
 this case as the standard model. We have made precollapse models by taking
 a density and internal energy distribution from the spherically
 symmetric 15 $M_\odot$ model by Woosley and Weaver (1995) and by adding
 the angular momentum according to the rotation laws stated above.
\subsection{Hydrodynamics}
The numerical method for hydrodynamic computations employed in this
paper is based on the ZEUS-2D code \cite{stone}.
The code is an Eulerian one based on the finite-difference method and
employs an artificial viscosity of von Neumann and Richtmyer to capture
shocks.  The self-gravity is managed by solving the Poisson equation
with the Incomplete Cholesky decomposition Conjugate Gradient (ICCG)
method. Spherical coordinates are used and one quadrant of the meridian
section is covered with 300 ($r$) $\times$ 50 ($\theta$) mesh
points. The code is checked by standard tests such as the Sod shock-tube
problem. We have made several major changes to the base code to include
the microphysics. First we have added an equation for electron fraction
to treat electron captures, which is solved separately. We have approximated electron
captures and neutrino transport by the so-called leakage scheme 
\cite{ep,blud,van1,van2}.
Second, we have incorporated the tabulated equation of state (EOS) based
on the relativistic mean field theory \cite{shen98} instead of the ideal
gas EOS assumed in the original code. It is noted that the
implementation of the recent realistic EOS to 2-D simulations is an
important progress beyond the previous calculations. 
 For a
more detailed description of the methods, see Kotake et
al. \cite{kotake}.
\subsection{Gravitational wave signal}
We follow the methods by \cite{mm,ys} in
order to compute the gravitational wave form. We will summarize it in the following for convenience. The amplitude of
the gravitational wave $h_{\mu \nu} \equiv g_{\mu \nu} - \eta_{\mu \nu}$ can be calculated by the quadrupole formula as follows:
\begin{equation}
h_{ij}^{\rm TT}(R) = \frac{2G}{c^4}\frac{1}{R}\frac{d^2}{dt^2}I_{ij}^{TT}(t-
\frac{R}{c}),
\label{quadru}
\end{equation}
where $i,j$ run from 1 to 3, $t$ is the time, $R$ is the distance
from the source to the observer, the superscript ``TT'' means to take
the transverse traceless part, and $I_{ij}$ is the reduced quadrupole
defined as 
\begin{equation}
I_{ij} = \int \rho(x) (x_i x_j - \frac{1}{3}x^2 \delta_{ij})\,d^3 x.
\end{equation}
 The second time derivative of the
reduced quadrupole formula, which is difficult to be treated
numerically, can be replaced by the spatial derivative by the equations
of motion. In the case of axisymmetric collapse, the transverse
traceless gravitational field is shown to have one independent
component, $h_{\theta \theta}^{\rm{TT}}$, and it is dependent solely on
$A_{20}^{\rm{E} 2}$. Then one derives for the component of $h^{\rm TT}$ the following formula,
\begin{equation}
h_{\theta \theta}^{\rm{TT}} = \frac{1}{8}\Bigl(\frac{15}{\pi}\Bigl)^{1/2} {\sin}^2 \theta \,\,\frac{A_{20}^{\rm{E} 2}} {R},
\end{equation}
where $\theta$ is the polar angle and $A_{20}^{\rm{E} 2}$ is defined
as
\begin{eqnarray}
 A_{20}^{\rm{E} 2} &=& \frac{G}{c^4} \frac{32 \pi^{3/2}}{\sqrt{15 \pi}} \int_{0}^{1} 
\int_{0}^{\infty}  r^2 \,dr \,d\mu \,\rho [ {v_r}^2 ( 3 \mu^2 -1) + {v_{\theta}}^2 ( 2 - 3 \mu^2)
 - {v_{\phi}}^{2} - 6 v_{r} v_{\theta} \,\mu \sqrt{1-\mu^2} \nonumber\\
 & & - r \partial_{r} \Phi (3 \mu^2 -1) + 3 \partial_{\theta} \Phi \,\mu
\sqrt{1-\mu^2}], 
\end{eqnarray}
where $ \partial_r = \partial/ \partial r,\,\,\partial_{\theta} =
\partial/ \partial \theta,\,\, \mu = \cos \theta$ and $\Phi$ is the
gravitational potential. 
Since the gravitational wave is radiated most in the equatorial plane,
the observer is assumed to be located in the plane in the following
discussions. In addition, the source is assumed to be located at a
distance of  $R = 10$ kpc. 
\section{Numerical Results}
\begin{table}
\caption{Summary of important quantities for all models. $t_{\rm b}$ is
 the time of bounce, $\rho_{\rm{max b}}$ is the maximum density at
 bounce, $M_{\rm i.c \,\, b} $ is the mass of inner core at bounce,
 $T/|W|_{\rm final}$ is the final ratio of rotational energy to
 gravitational energy of the core,
 $\Delta t$ is the duration time (FWHM) of the first burst, and $|h^{\rm{TT}}|_{\rm max}$ is the maximum amplitude of the first
 burst. Note that we speak of the inner core, where the matter falls
 subsonically, which corresponds to the unshocked region after core bounce. } 
\label{table:2}
\begin{center}
\begin{tabular}{ccccccc} \hline \hline
Model & $t_{\rm b}$ & $\rho_{\rm{max b}}$  & $M_{\rm i.c \, b}$ 
 & $T/|W|_{\rm final} $ & $\Delta t$   & $|h^{\rm{TT}}|_{\rm max}$ \\
      & (ms)   & ($10^{14}\,\,\rm{g}\,\,\rm{cm}^{-3}$) & $(M_{\odot})$ &
 $(\% )$&(ms)
 & $(10^{-20})$ \\ \hline
 SSL & 227.4   & 2.95    & 0.74   & 4.87  & 0.58   &  1.03     \\ 
 SSS & 227.3   & 2.80    & 0.75   & 5.66  & 0.54   &  1.49     \\ 
 SCL & 241.2   & 2.87    & 0.75   & 5.09  & 0.46   &  0.93\\ 
 SCS & 230.7   & 2.84    & 0.73   & 6.46  & 0.56   &  2.00 \\    
 MSL & 242.7   & 2.65    & 0.75   & 8.44  & 0.72   &  1.58 \\
 MSS & 242.2   & 2.15    & 0.87   & 9.05  & 0.57   &  2.03 \\
 MCL & 241.9   & 2.85    & 0.76   & 8.38  & 0.70   &  1.53 \\  
 MCS & 245.4   & 1.45    & 0.91   & 9.94  & 0.51   &  2.87\\
 RSL & 338.7   & 1.21    & 0.94   & 14.6  & 2.89   &  0.48\\
 RSS & 328.6   & 0.53    & 0.96   & 13.3  & 2.80   &  0.78     \\
 RCL & 338.4   & 1.36    & 0.95   & 14.5  & 4.47   &  0.42     \\
 RCS & 326.7   & 0.18    & 1.11   & 12.3  & 1.91   &  1.41      \\\hline \hline
\end{tabular}
\end{center}
\end{table}

\subsubsection{The Properties of the Waveform}
We first show the general properties of the waveform with collapse dynamics. 
 We choose model MSL (standard) as a representative model. For later 
convenience, the values of several important quantities are
summarized in Table \ref{table:2}. The time
evolution of the amplitude of gravitational wave and the central density near core bounce are shown in Figure \ref{MSL}. 
As the inner core shrinks, the central density increases and a core
bounce occurs when the central density reaches its peak at $t_{\rm b} \cong
243 ~\rm{msec}$ with $2.65 ~10^{14} ~\rm{g}~\rm{cm}^{-3}$. At this time,
the absolute value of the amplitude becomes maximum. After the core
bounce, the core slightly re-expands and oscillates around its equilibrium.  
As a result, the gravitational wave shows several small bursts and
begins to decay. These gross properties are common to all other
models. However, there exist some important differences when we compare
them in more detail. We will discuss the differences in the following.

In Figure \ref{MCS}, the time evolution of the amplitude of
gravitational wave and the central density near core bounce 
for model MCS are given. For the model, it is noted that the initial
rotation law is cylindrical with strong differential rotation.
By comparing the left panel of Figure \ref{MSL} to \ref{MCS}, the
oscillation period of the inner core for model MCS is clearly longer
than for model MSL. In other words, the pronounced peaks can be seen
 distinctively in this case. This is because the central density becomes
 more smaller after the distinct bursts (compare the right panels of Figure 
\ref{MSL} to \ref{MCS}). This effect increases with the initial angular
momentum (compare the left panel of Figure \ref{MCS} to \ref{mix}).
 It is also found that the signs of the values of the second peaks are
 negative for model MCS, on the other hand, positive for model MSL 
 (compare the left panels of Figure \ref{MCS} to \ref{MSL}). Note that we will speak of the second peak
where the absolute amplitude is second largest. 
The above characteristics are
common to the models for strongly differential rotation with
cylindrical rotation law (see Table {\ref{table:3}). A specific feature
in the waveform is found for models RSL and RSS in which there exist the
small peak or shoulder before the peak burst (see the right panel of
Figure \ref{mix}).

Next we will compare our results of the waveforms with those by Zwerger et
al. \cite{zweg} who categorized the shapes of waveforms into three
district classes. They used a polytropic EOS to express the pressure from 
the degenerate leptons as $P \propto \rho^{\gamma}$. They reduced the 
adiabatic index,
$\gamma$, from 1.325 to 1.28  below the nuclear density regime in order to 
approximate the related microphysics and neutrino transport.
They assumed the cylindrical rotation law for all the models and varied
both the degree of
differential rotation and the initial angular momentum. 
Our models except for the strongly differential
rotation with cylindrical rotation law correspond to so-called the type
I in their nomenclature. On the other hand, our models 
for strongly differential rotation with cylindrical (not shellular)
rotation law correspond to type II. 
It should be noted that type II signals were limited to occur only for
rapid, strongly differential rotation with cylindrical rotation law in
their models. Furthermore in this study, we find that type II does not
occur for models with shellular rotation law regardless of the degree of
differential rotation and the initial rotation rate, on the other hand, does occur regardless of
the initial rotation rate in case of strong differential rotation
with cylindrical rotation law. Finally type III 
occurred only when the core collapses very rapidly  ($\gamma = 1.28$) in their
calculations. There are no models
which correspond to the type in our calculations. This is because our realistic
EOS is not so soft in the corresponding density regime.

\begin{table}
\caption{Some characteristic quantities for the waveform analysis. The names of
 the models whose initial rotation law is cylindrical with strong
 differential rotation are written in bold letters.
$T^{I}_{\rm osc}$ and $T^{II}_{\rm osc}$ the first and second
 oscillation period of the inner core, respectively.  $h^{TT}_{\rm second}$
 is the amplitude of gravitational wave at the second peak.}
\label{table:3}
\begin{center}
\begin{tabular}{cccc} \hline \hline
Model & $T^{I}_{\rm osc}$   & $T^{II}_{\rm osc}$&  $h^{TT}_{\rm second}$ \\
     & [ms]       & [ms]   & $[10^{-20}]$ \\\hline
 SSL & 1.6        & 2.4       & 0.82      \\ 
 SSS & 1.5        & 2.2       & 0.94 \\ 
 SCL & 1.6        & 2.3       & 0.75  \\ 
 \bf{SCS} & 1.8      & 2.6       & -0.57           \\    
 MSL &  1.8          & 1.2       & 1.23  \\
 MSS &  2.2          & 2.4       & 0.98  \\
 MCL &  1.7          & 1.1       & 1.20    \\  
 \bf{MCS}& 3.1        & 2.6       & -0.79     \\
 RSL &  1.0          & 3.1       & 0.32 \\
 RSS &  1.0          & 7.8       & 0.20  \\
 RCL &  9.0          & 2.9       & 0.25       \\
 \bf{RCS} & 10.7     & 8.6       & -0.49  \\\hline \hline
\end{tabular}
\end{center}
\end{table} 
\subsubsection{Maximum Amplitude and Second Peak}
We will first discuss the relation between the maximum amplitudes of
gravitational wave and the initial $T/|W|$. The maximum
values and the related quantities are given in Table \ref{table:1} and 
\ref{table:2}.
From Figure \ref{max_hikaku}, it is found that the amplitude is largest
for moderate initial rotation rate (i.e., $T/|W|_{\rm init} = 0.5
\%$) when one fixes an initial rotation law and a degree of differential
rotation. This is understood as follows. The amplitude of gravitational
wave is roughly proportional to the inverse square of the typical
dynamical scale, $t_{\rm dyn}$ (e.g., Eq. (\ref{quadru})). Since  
$t_{\rm dyn}$ is proportional to the inverse root of the central
density $\rho$, the amplitude is proportional to the density. As a
result, the amplitude becomes smaller as the initial rotation rates
become larger because the density decreases for large initial rotation
rate (see Table \ref{table:2} and Figure \ref{max_rho_hikaku}). 
On the other hand, the amplitude is proportional to the 
value of the quadrupole moment, which becomes large in turn as the
total angular momentum increases. This is because the stronger
centrifugal forces make not only the mass of the inner core larger (see
Table \ref{table:2}), but
also deforms the inner core. The amplitude is determined by the
competition of these factors. As a result, the amplitude becomes
extrema for moderate initial rotation rates in our calculation.

Next we will discuss the values of maximum amplitude between our results
and those by other groups. The values of maximum amplitude for all our
models range from $5\times 10^{-21} \leq h^{\rm TT} \leq 3\times 10^{-20}$  
, which is almost the same as the results by Zwerger et al. \cite{zweg}
and M\"{o}nchmeyer et al. \cite{mm}. On the other
hand, the values of the standard models by Yamada et al. \cite{ys} are about an order of
magnitude lower. This is understood as follows. 
Yamada et al. \cite{ys} employed a parametric EOS by which the effects
of microphysical and transport processes were assumed to be
expressed. Since they found that the maximum amplitude
is most sensitive to the adiabatic index at the subnuclear density by
their parametric surveys, we pay attention to this. 
The comparison of the effective adiabatic index between our EOS and their EOS is shown in 
Figure \ref{gamma_hikaku}.
Compared to their EOS, our realistic EOS is rather soft for the
subnuclear density regime. At the regime, our realistic EOS can express
the softening of EOS by the effect that the nuclear interactions become
attractive.  Due to this effect, the inner core can shrink more compact at core
bounce, which results in the larger maximum amplitude. 
It is naturally suggested that we may get the information about
the subnuclear matter if we can detect the gravitational wave from
the rotationally collapsing cores.  We hope it can be realized in the
near future since 
the maximum amplitudes for our models are mostly  
within the detection limit for TAMA and first LIGO
which are now in operation if a source is located at a distance of 10
kpc (see
Figure \ref{detect}).
   
As pointed earlier, we find that the signs of the values of the second peaks are negative for the models for strongly differential rotation with
cylindrical rotation law and positive for the others 
(see Table {\ref{table:3}). The absolute amplitudes of the second peak
are also presented in Figure \ref{detect}. As
 shown, they are within the detection
 limit of first LIGO for a source at a distance of 10 kpc. In addition, it seems quite possible 
for the detectors in the next generation such as advanced LIGO and LCGT
to detect the difference. 
Therefore if we can find the difference of the signs of the second peaks by observations of gravitational
wave, we will obtain the information about the angular momentum distribution of evolved massive stars. Since there is no way except for the detection
of the gravitational wave to obtain the information, it is of great
importance to detect the second peaks in the future.   
\subsubsection{Secular instability}
In some models (Models RSL and RCS) with large initial rotation rate,
the final rotation rate exceeded the critical value (see Table
\ref{table:2}), where MacLaurin spheroids become secularly unstable against tri-axial perturbations
($T/|W| > T/|W|_{\rm seq} = 13.75 \% $). 
Rampp et al. \cite{rampp} reported that no considerable enhancement of
gravitational radiation due to the growth of secular instabilities was
found within a time scale of several 10 ms after core bounce. 
Since our calculations are within the time scale, the axial symmetry assumed
in this work may be justified.  
 \section{Conclusion}
We have done a series of two-dimensional hydrodynamic simulations of the
rotational collapse of a supernova core and calculated gravitational
waveforms using the quadrupole formula. We have employed a realistic EOS
and taken into account electron captures and neutrino transport in an
approximated method. We have found the following:

1. The peak amplitudes of gravitational wave obtained in this study 
 are mostly within the detection limits of the 
detectors of TAMA and first LIGO which are now in operation if a source is
 located at a distance of 10 kpc. In addition, the peak amplitude becomes extrema for the models whose initial rotation rate is moderate.

2. The waveforms are categorized into the criteria by Zwerger et
al. \cite{zweg}. In addition,
we further find that type II does not
occur for models with shellular rotation law, on the other hand, 
does occur regardless of
the initial rotation rate in case of strong differential rotation
with cylindrical rotation law, and that the type III does not occur if a
realistic EOS is employed.

3. At the subnuclear density regime, our realistic EOS can express the
softening of EOS by the effect that the nuclear interactions become
attractive. Therefore the inner core can shrink more compact at core
bounce than the other work \cite{ys} in which EOS is expressed in a parametric
manner. Subsequently, this results in the larger maximum amplitude. 
It follows that we may get the information about
the subnuclear matter if we can detect the gravitational wave from
the rotationally collapsing cores.

4. The signs of the values of the second peaks are negative for the models with strong differential
rotation with the cylindrical rotation law, on the contrary, positive
for the other models. The absolute amplitudes of 
the second peaks are within the detection limit of first LIGO. Therefore if we 
can detect the signs of the second peaks, it will give us
the information as to the angular momentum distribution of massive
evolved stars when a supernova occurs at our galactic center.

As stated earlier, the detection of gravitational wave is likely for the models whose initial rotation rate is moderate.
According to the study of rotational core collapse by Kotake et
al. \cite{kotake}, the anisotropic neutrino radiation is well induced by such a 
rotation rate. Noting that the anisotropic neutrino radiation can
induce the globally asymmetric explosion \cite{shimi01}, the detection of
gravitational wave will
be a good tool to help understand the explosion mechanism itself. 

\section*{Acknowledgments}
We are grateful to K. Numata and M. Ando for helpful discussions.
K.K would like to be thankful to M. Oguri and M. Shimizu for supporting
computer environments. 
The numerical calculations were partially done on the
supercomputers in RIKEN and KEK (KEK supercomputer Projects No.02-87 and
No.03-92). This work was partially supported by 
Grants-in-Aid for the Scientific Research from the Ministry of
Education, Science and Culture of Japan through No.S 14102004, No.
14079202, and No. 14740166.

\clearpage
\begin{figure}[hbt]
\begin{center}
\includegraphics[scale=0.64]{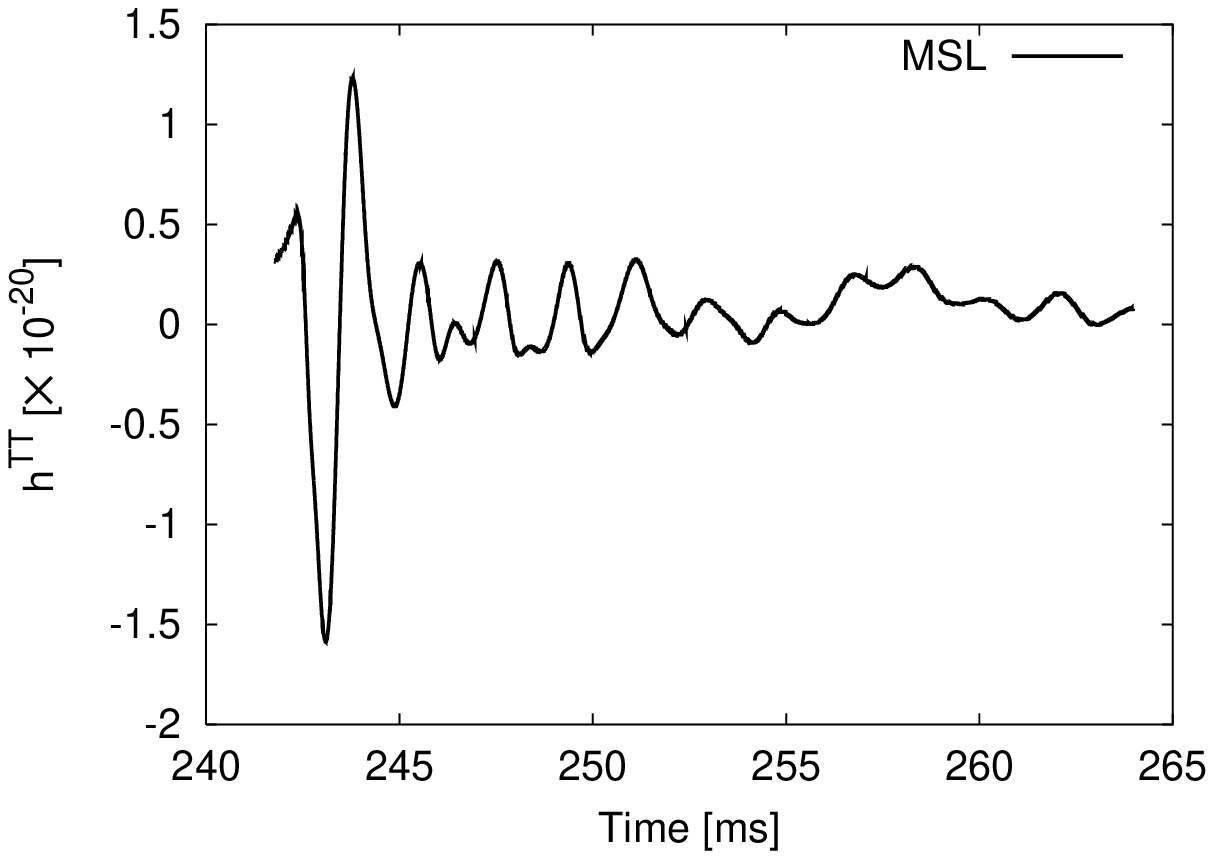}
\includegraphics[scale=0.64]{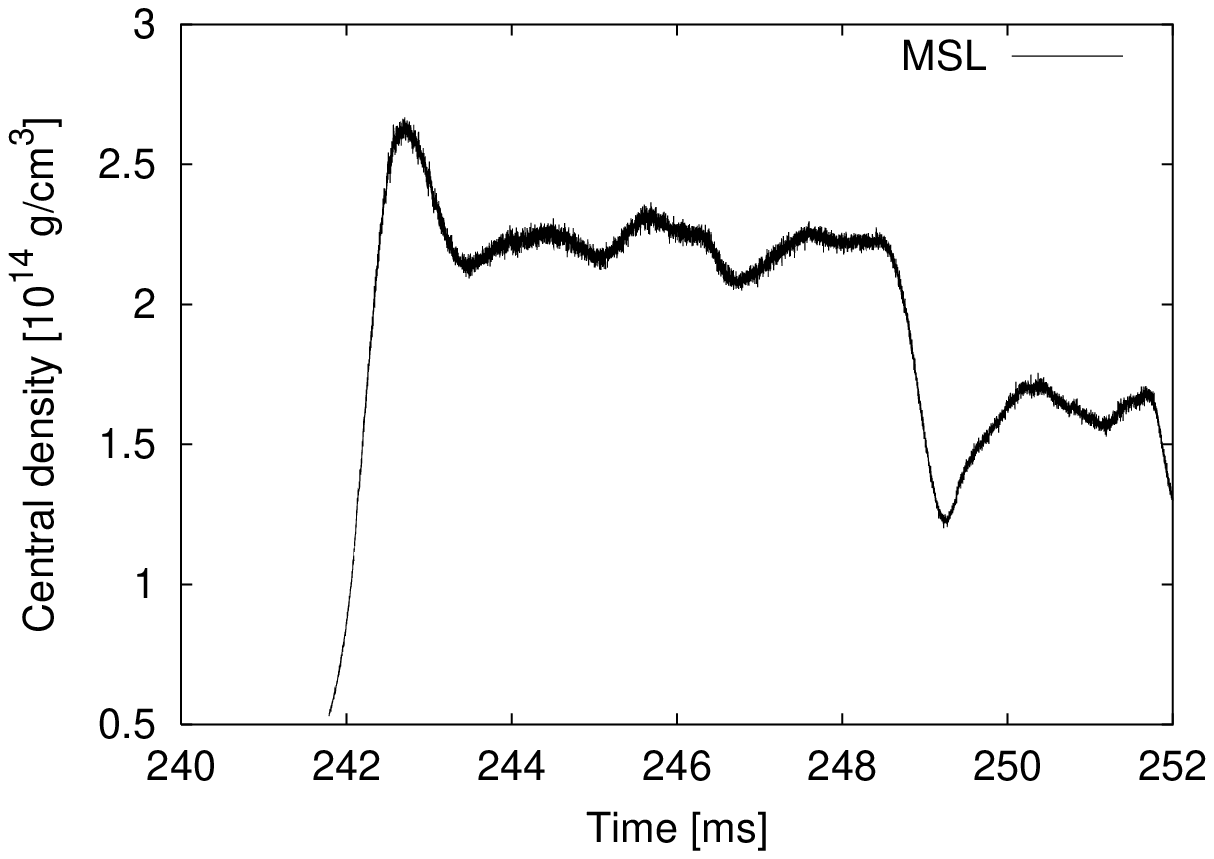}
\vspace{0.5 cm}
\caption{Time evolution of the amplitude of gravitational wave (left
 panel) and central density near core bounce (right panel) for model MSL.}
\label{MSL}
\end{center}
\end{figure}

\begin{figure}[hbt]
\begin{center}
\includegraphics[scale = 0.64]{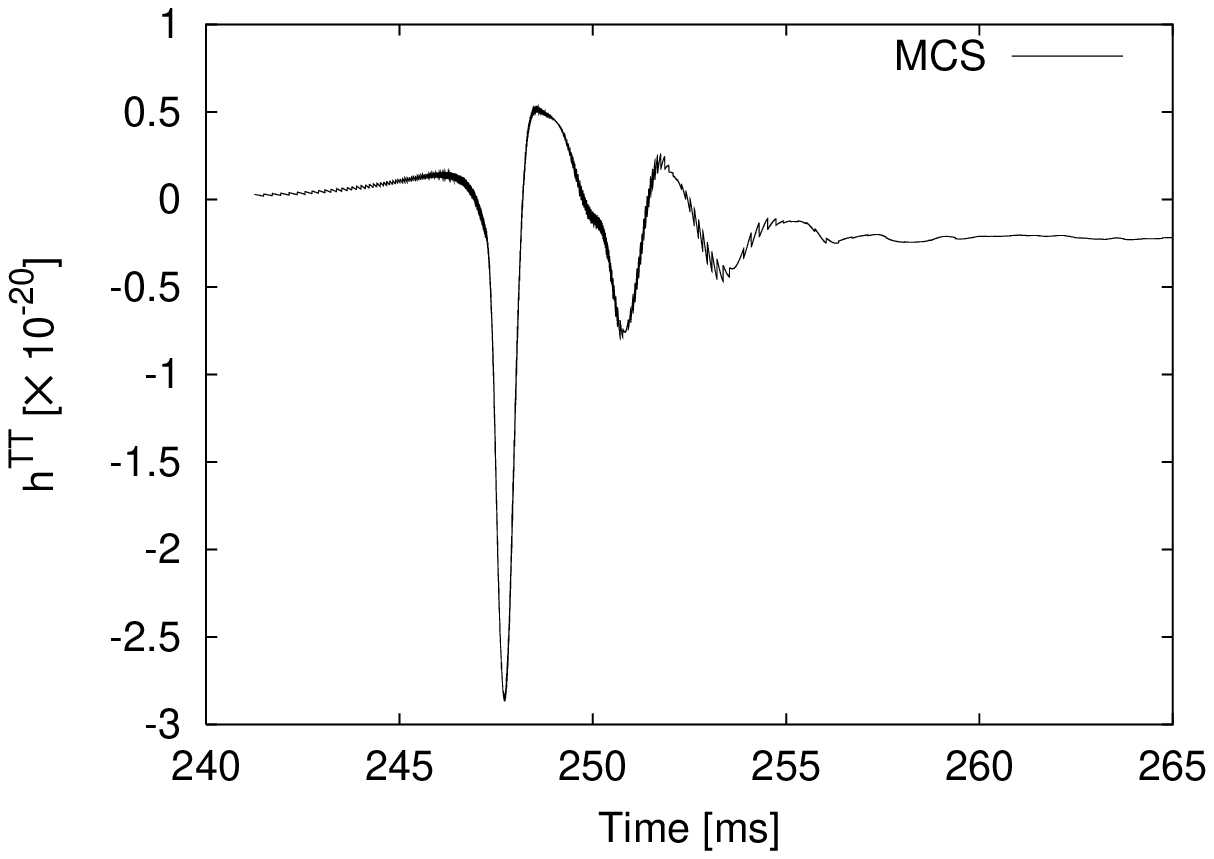}
\includegraphics[scale = 0.64]{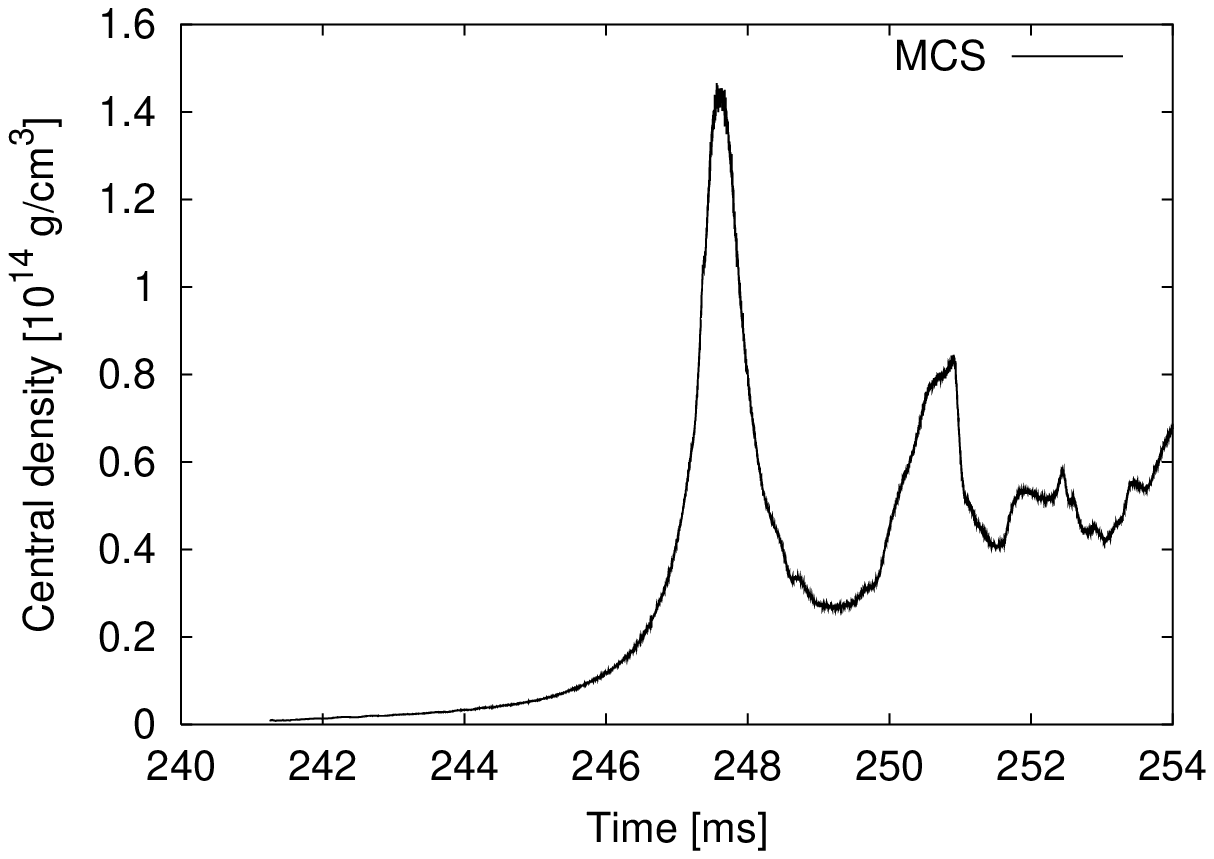}
\caption{Same as Figure \ref{MSL} but for model MCS.}
\label{MCS}
\end{center}
\end{figure}

\begin{figure}[hbt]
\begin{center}
\includegraphics[scale = 0.64]{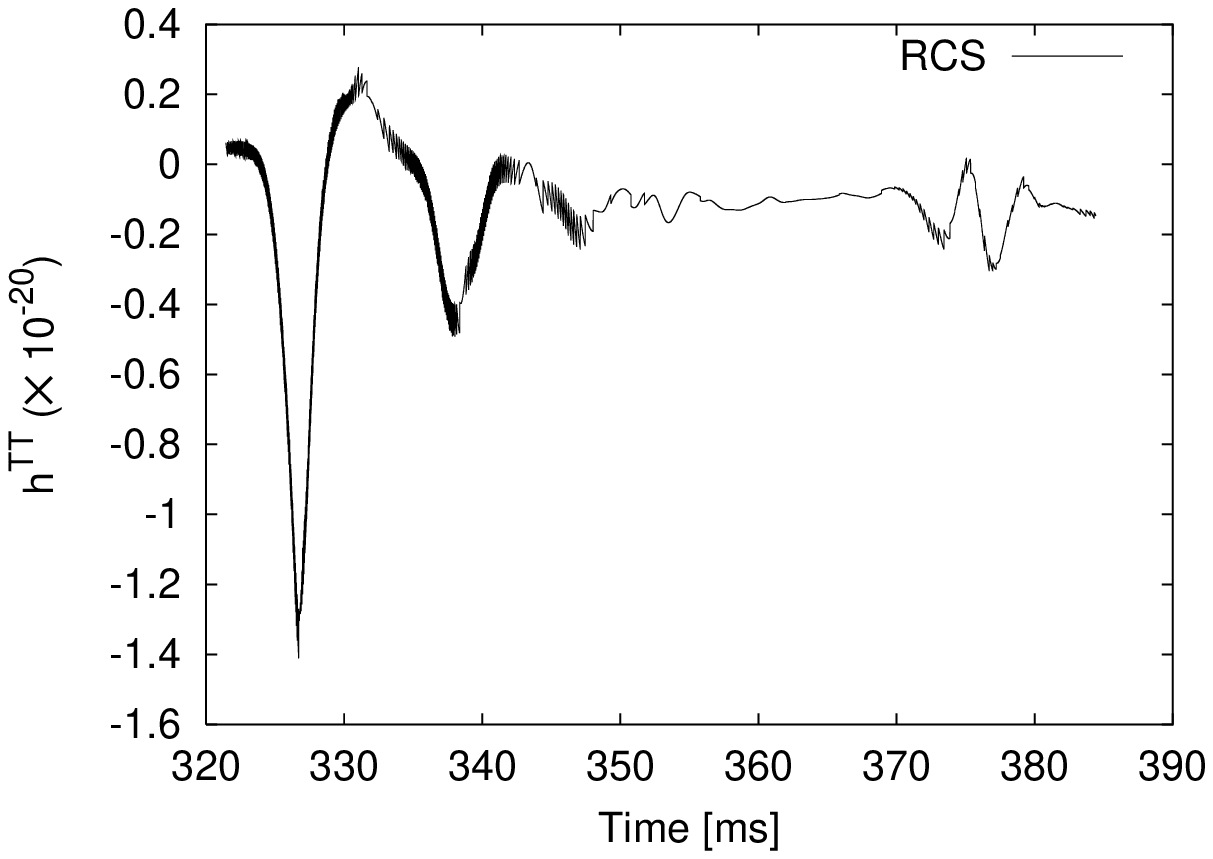}
\includegraphics[scale = 0.64]{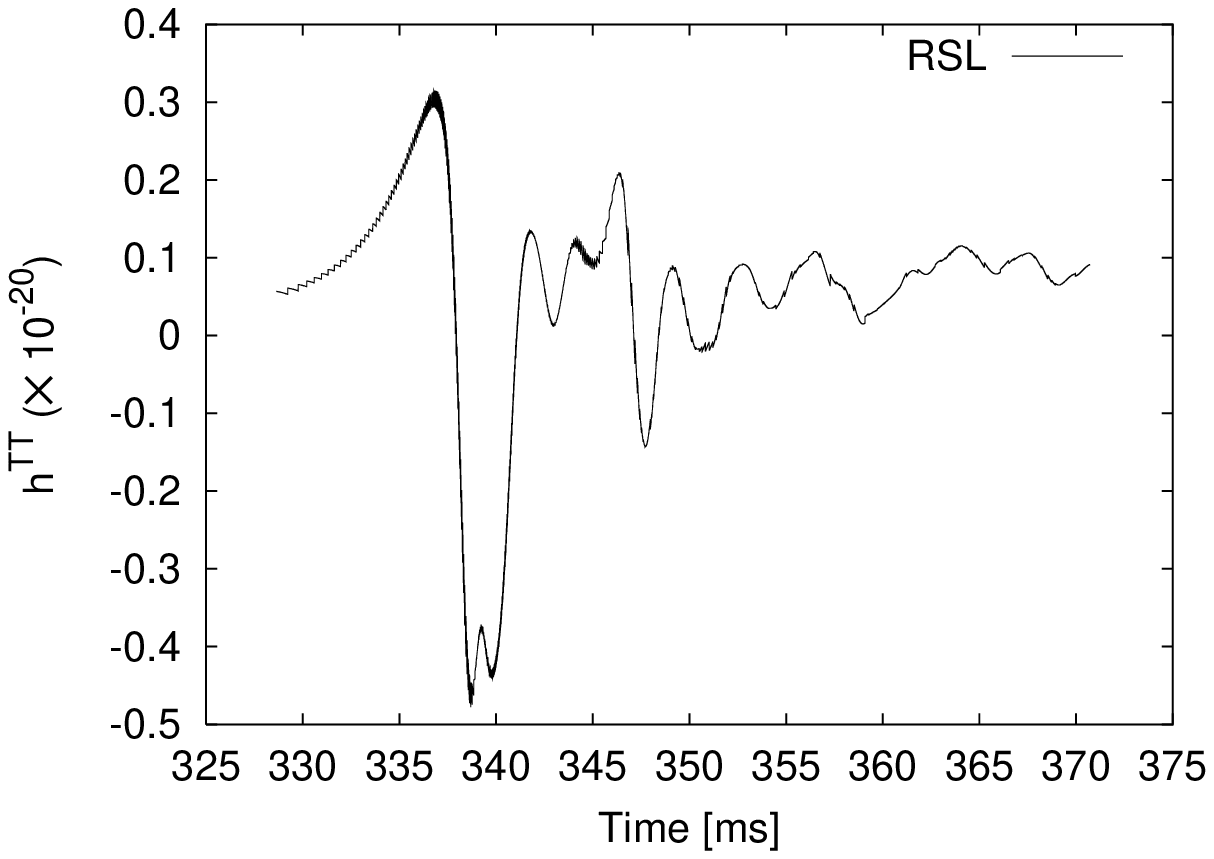}
\vspace{0.5 cm}
\caption{Waveforms for models RCS (left panel) and RSL (right
 panel). For the left panel, very district peaks are shown. For the right panel, a small peak and shoulder after the first burst
 can be seen.}
\label{mix}
\end{center}
\end{figure}

\begin{figure}[htb]
\includegraphics[scale = 0.8]{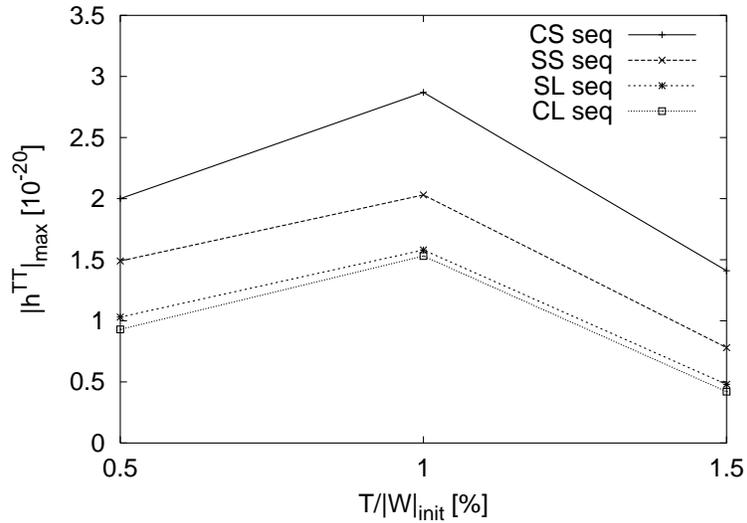}
\caption{Relation between $T/|W|_{\rm init}$ and the peak amplitude
 $|h^{\rm TT}|_{\rm max}$ for all the models. 
``CS, SS, SL, CL seq'' in the figure represent the model sequences whose
 names are taken from the second and third letters in Table
 \ref{table:1}, respectively. Note that the second and third letters
 mean the rotation law and the degree of differential rotation,
 respectively. }
\label{max_hikaku}
\end{figure}
\begin{figure}
\includegraphics[scale = 0.8]{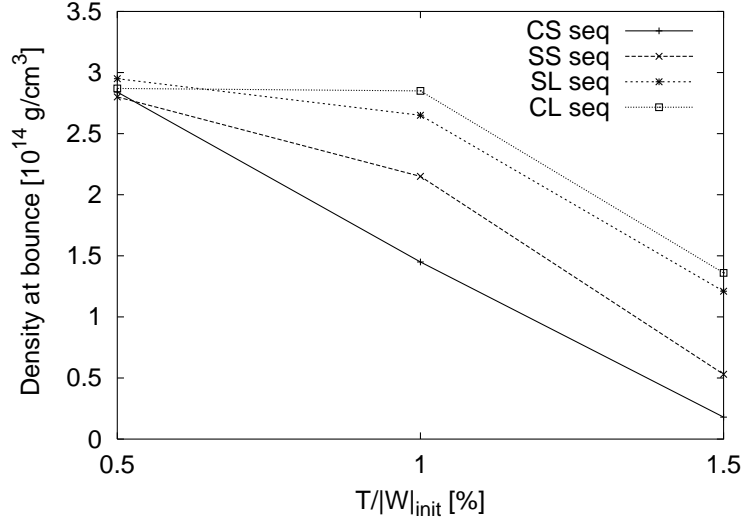}
\caption{Relation between $T/|W|_{\rm init}$ and the central density at core bounce
 $(10^{14}\,\rm{g}\,\,\rm{cm}^{-3})$ for all the models. The meaning of
 the label of lines is the same as Figure \ref{max_hikaku}. It is found
 that the central density at the core bounce becomes smaller as the
 initial angular momentum becomes large.}
\label{max_rho_hikaku}
\end{figure}

\begin{figure}
\includegraphics[scale = 0.8]{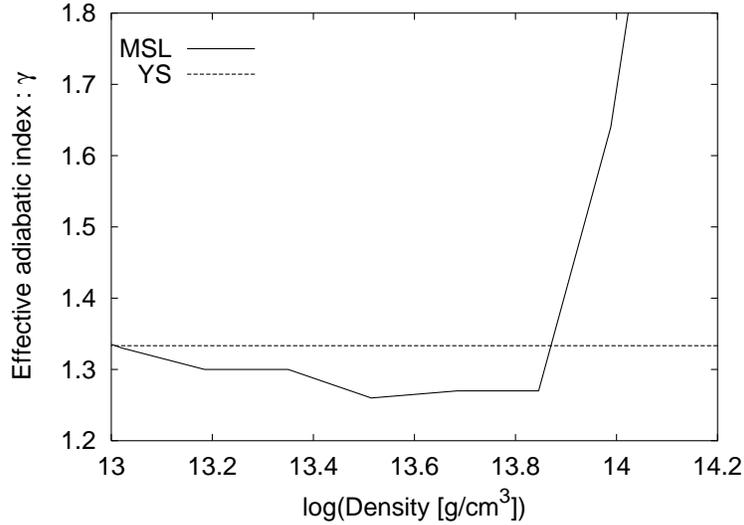}
\caption{Relation between the density and the effective adiabatic 
 index $\gamma$ at the subnuclear density. MSL represents the relation
 for our standard model. On the other hand, the index employed in the
 study of Yamada et al. \cite{ys} is represented as YS.}
\label{gamma_hikaku}
\end{figure}
\begin{figure}
\includegraphics{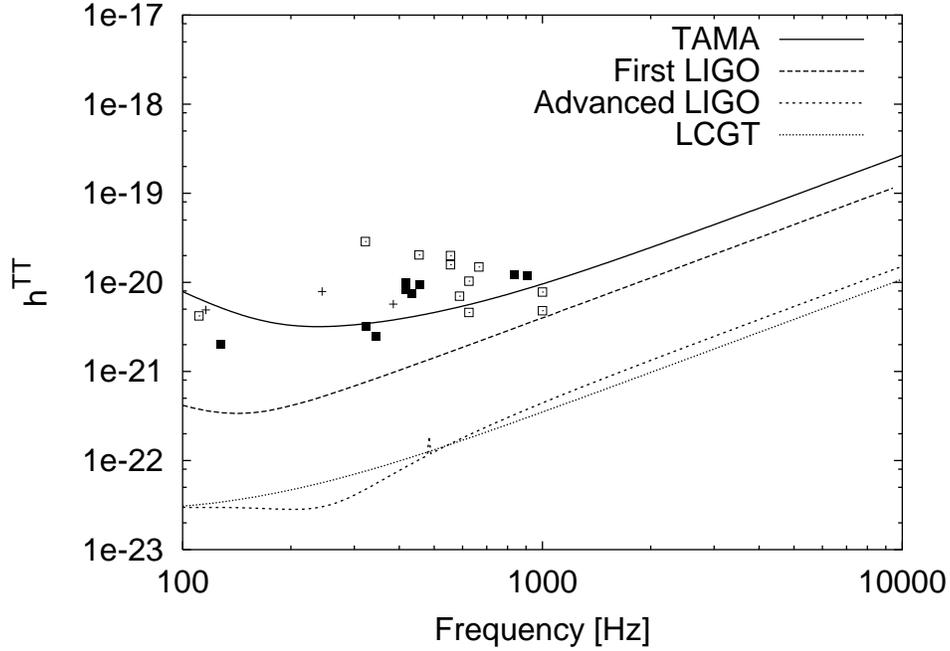}
\caption{Detection limits of TAMA \cite{tama}, first LIGO
 \cite{firstligo}, advanced LIGO \cite{advancedligo}, and LCGT \cite{lcgt} with
 the expected amplitudes from numerical simulations. The open squares
 represent the maximum amplitudes for all the models. On the other hand,
 the pluses and the closed squares represent the amplitudes of second
 peaks for models with strong differential rotation with cylindrical
 rotation law and for the other models, respectively. We estimate the characteristic
 frequencies by the inverse of duration periods of the corresponding 
 peaks. 
Note that the source is assumed
 to be located at the distance of 10 kpc.}
\label{detect}
\end{figure}
\end{document}